\documentclass[9pt,twocolumn,twoside]{osa-article}

\journal{osajournal}

\articletype{Research Article}

\usepackage{float}
\usepackage{graphicx}
\usepackage{dcolumn}
\usepackage{bm}
\usepackage{hyperref}
\usepackage{siunitx}
\usepackage{setspace}
\DeclareMathOperator\erf{erf}
\renewcommand{\copyright}{\textsuperscript{\tiny{\textcopyright}}}

\begin{document}

\title{Optimization of diamond optomechanical crystal cavities}

\author{Flávio Moraes,\authormark{1,2} Gabriel H. M. de Aguiar,\authormark{1,2} Emerson G. de Melo,\authormark{3} Gustavo S. Wiederhecker,\authormark{1,4} and Thiago P. Mayer Alegre~\authormark{1,2*}}

\address{
\authormark{1}Photonics Research Center, University of Campinas, Campinas, SP, Brazil\\
\authormark{2}Applied Physics Department, Gleb Wataghin Physics Institute, University of Campinas, Campinas, SP, Brazil\\
\authormark{3}Department of Materials Engineering, Lorena School of Engineering, University of S\~ao Paulo, Lorena, SP, Brazil\\
\authormark{4}Quantum Electronics Department, Gleb Wataghin Physics Institute, University of Campinas, Campinas, SP, Brazil\\
}

\email{\authormark{*}alegre@unicamp.br} 


\begin{abstract}
Due to recent development of growing and processing techniques for high-quality single crystal diamond, the large scale production of diamond optomechanical crystal cavities becomes feasible, enabling optomechanical devices that can operate in higher mechanical frequencies and be coupled to two-level systems based on diamond color centers. In this paper we describe a design optimization method to produce diamond optomechanical crystal (OMC) cavities operating at the high-cooperativity regime (close to unity) at room temperature.
\end{abstract}

\section{Introduction}
Optomechanical crystal (OMC) cavities are micro- or nanofabricated structures that take advantage of the coupling between optical and acoustic waves to generate superposition between mechanical and optical cavity modes~\cite{Eichenfield2009, Aspelmeyer2014}, allowing amplification, generation and detection of photons and phonons~\cite{Kippenberg2005, Arcizet2006, Chan2011, Weitz2013, Teufel2009, Galland2014}. Most applications of OMC cavities rely on the capability of laser cooling the mechanical resonator to its ground state~\cite{Chan2011}, where a true quantum-coherent interactions may take place. In principle, ground state cooling may be achieved even at room temperature, by strongly pumping the cavity with a red-detuned laser. In such a scenario the optical cavity suppress Stokes scattering while amplifying the anti-Stokes scattering, which effectively drains energy from the mechanical resonator. The OMC laser cooling is efficient if the optomechanical coupling rate ($g_0$) is large enough to overcome the photons and phonon dissipation rates. This condition is commonly referred to as high-cooperativity regime ($C=4n_pg_0^2/(\kappa \Gamma)>1$)~\cite{Aspelmeyer2014}, where $n_p$ is the number of pump photons inside the cavity and $\kappa$ and $\Gamma$ are the cavity optical and mechanical decay rates. Another condition for reaching efficient cooling is related to the capability of the optical cavity to distinguish between Stokes and anti-Stokes, referred to as the sideband-resolved regime ($\omega_m\gg\kappa$), where $\omega_m$ is the resonance frequency of the mechanical oscillator.

Diamond OMC cavities are specially interesting for optomechanics due to the remarkable mechanical and optical properties of diamond~\cite{Burek2016, Chia2021}. For instance, the large bandgap of diamond, $\SI{5.45}{\eV}$, combined with its high thermal conductivity and small thermal expansion coefficient, allows fabrication of nanobeams operating with large number of pump photons ($n_c>10^4$)~\cite{Burek2016}. At the same time the large Young's modulus of diamond, combined with its low intrinsic mechanical dissipation (e.g., low thermoelastic damping) enables the fabrication of mechanical resonators with high resonance frequencies and high quality factors ($Q_m > 10^5$)~\cite{Tao2014}. At last, diamond also have been seen as a promising material for quantum networks since it can host resilient single-spin color centers~\cite{Gruber1997} to store quantum information~\cite{Hausmann2013, Burkard2014}. OMC cavities can be used to manipulate diamond color centers with light at telecom wavelengths, without relying on qubit optical transitions~\cite{Shandilya2021}.

Diamond devices are however not simple to process~\cite{Castelletto2017}.  Standard planar fabrication techniques do not work for bulk diamond~\cite{Piracha2016} and single crystal diamond (SCD) grown by Chemical Vapor Deposition (CVD) are not uniform over large areas~\cite{Burek2012}. SCD usually needs to be built with a relatively larger thickness and then thinned to the device thickness~\cite{Lee2012}. Since only the upper surface of the SCD has good optical quality, it must be either thinned from bottom to up with quasi-isotropic etching, mounted upside down over another material through wafer bonding, or flipped using electrochemical lift-off~\cite{Khanaliloo2015}. Resonators in mono-crystalline CVD grown SCD can be produced by removing the lower-quality nucleation layer with high-energy ion implantation, leaving the cavity in the highest quality layer at the diamond surface~\cite{Riedrich-Moller2012}. A quasi-isotropic etch technique has been applied to undercut the resonator structure achieving optical quality factor above $1.4\times10^4$~\cite{Mouradian2017}. The optical quality factor in these cases is affected by the residual damage of the etching in the structure. Resonators in polycrystalline diamond (PCD) thin films, in contrast, are much simpler and cheaper to process since they can be grown with the resonator thickness. However, they have limited optical quality factors due to absorption and light scattering across grain boundaries ($Q_o\sim10^5$)~\cite{Abdou2018}.

In this work, we are interested in developing an optimized PCD-based nanobeam design capable of operating at both the high-cooperativity and the resolved-sideband regime. These regimes have already been reached in SCD-based microdisks~\cite{Mitchell2016} and triangular nanobeams~\cite{Burek2016}. PCD-based device design optimization yet has to consider the limited optical quality factor of the material, using strategies to boost cavity performance~\cite{Babinec2011} while aiming in geometries with high optomechanical coupling to increase the cooperativity. A widely spread design optimization method developed for Si-based nanobeams~\cite{Chan:2012iy} focuses on maximizing $Q_og_0^2$ by combining finite-element (FEM) simulations with numerical optimizations methods. When applied to diamond devices, however, they lead to designs with a low mechanical quality factor~\cite{Cady2019}. Because diamond has a lower refraction index ($n=2.45$) and a higher Young's modulus than silicon, it has a weaker localization of the optical cavity mode and a stronger localization of the mechanical mode. As consequence, to increase the overlap between optical and mechanical modes, numerical methods tend to converge to geometries with low confinement of mechanical modes. Under these circumstances, we had to develop a new optimization method considering optical and mechanical losses. Some advantages of our method are: (i) it leads to a better solution after a relatively small number of iterations due to the hybrid optimization process that starts with a genetic algorithm (GA) followed by a pattern search (PS) for local optimization; (ii) it returns a set of locally optimized solutions, instead of a single solution; (iii) it restricts solutions to a specific-frequency range, instead of fixing geometric parameters; (iv) our output geometries already have the resonant optical mode in the optical communication wavelength bands and don't need to be rescaled.

We present here, two distinct robust geometries with high single-photon cooperativity ($C_0>10^{-2}$) for diamond nanobeams, that were delivered by our optimization method.

\section{Cavity design}

\begin{figure*}[ht!]
\begin{center}
\includegraphics[width=12.5cm]{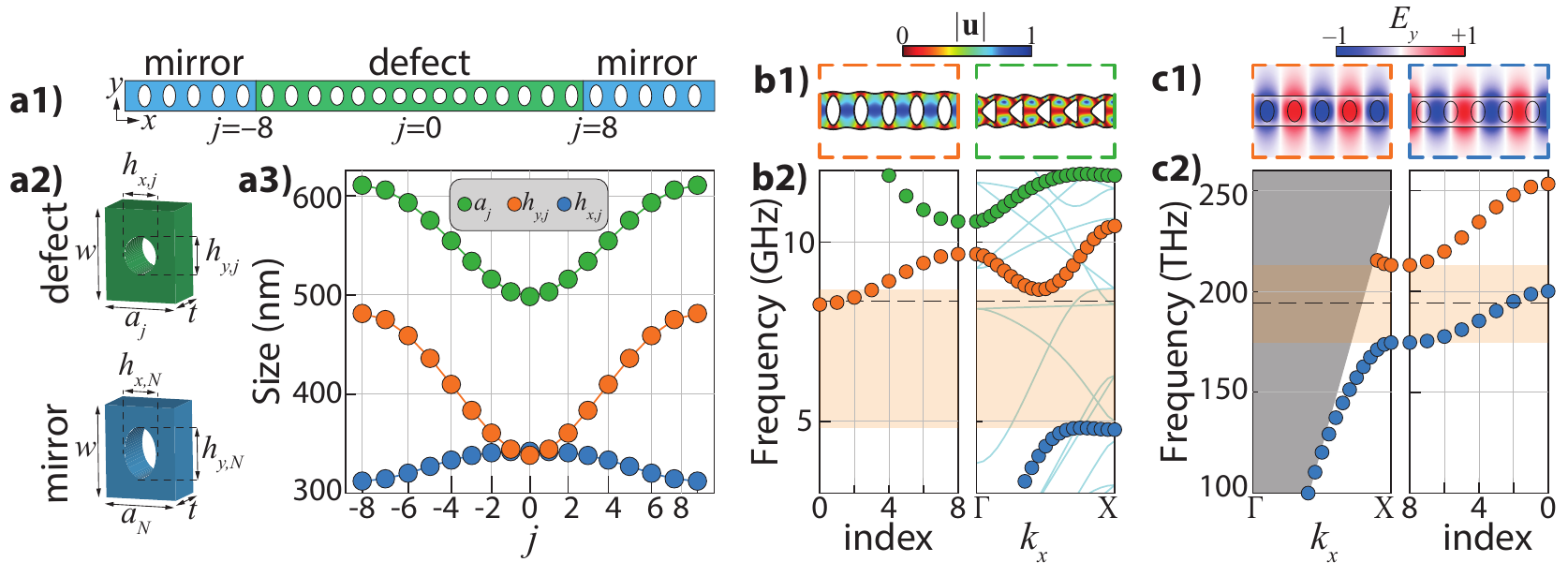}
\caption{\small{\textbf{a1)} Representation of the rectangular nanobeam pattered with quasi-periodic elliptical holes~\cite{Chan2012}. \textbf{a2)} Ilustration of the central defect ($j=0$) and mirror mirror (j=8) nominal unit cells, with thickness $t$, width $w$, lattice parameter $a_j$ and hole diameters $h_{xj}$ and $h_{yj}$. \textbf{a3)} Defect cells parameters variation for an optimized solution of the nanobeam.
\textbf{b)} Mechanical band diagram~(\textbf{b2}-right) and the variation of the unit cell mechanical frequency at the $\Gamma$-point for each defect from the center ($j=0$) to the mirrors ($j=8$) (\textbf{b2}-left). The blue, orange and green dots represent the zero-, first-~(\textbf{b1}-left) and second-order~(\textbf{b1}-right) symmetric modes, respectively. The zero-order mode (blue dots) at the $\Gamma$-point comes down to a translation of the unit cell. The continuous blue lines represent the non-symmetric modes and the transparent red area represents the bandgap at the mirror. The parameters used for the defects and mirrors (\textbf{a3}) were obtained from a solution of the nanobeam's optimization, and the dashed line represents the frequency of the confined optical mode for such nanobeam.
\textbf{c)} Optical band diagram~(\textbf{c2}-left) and the variation of the unit cell optical frequency at the X-point for each defect from the mirror ($j=8$) to the center ($j=0$)~(\textbf{c2}-right). The blues and orange dots represent the zero-~(\textbf{c1}-right) and first-order~(\textbf{c1}-left) modes. The gray area represents a continuum of radiation. The dashed line represents the frequency of the lower-energy optical mode confined at the center of the optimized nanobeam.}}
\label{fig:nanobeam}
\end{center}
\end{figure*}

Our optimized cavity is a nanobeam with rectangular cross-section (width $w=\SI{664}{\nm}$ and thickness $t=\SI{450}{\nm}$)  patterned with elliptical holes (Fig.~\ref{fig:nanobeam}\textbf{a}). The nanobeam design consists of a sequence of unit cells (rectangular blocks with an oval hole in the middle, as depicted in Fig.~\ref{fig:nanobeam}\textbf{b}). From the central cell to any of the borders, there are $13$ unit cells. The first $N=8$ are defects cells followed by 5 mirror cells. 

For our nanobeam, the design optimization boils down to defining, in a way to maximize the cooperativity, the unit cell area (cell width and lattice parameter) and its hole size and shape, for the mirrors and the central defect cells, together with how they change along the nanobeam. The mirrors act as a waveguide with defined propagating modes and a bandgap between them. The defects are fabricated in a way that the energy of a specific mode deepens into the mirror bandgap, confining it locally. Figs.~\ref{fig:nanobeam}~\textbf{b} and \textbf{c} show the mechanical and optical band diagrams for propagating waves in a unit cell with Floquet periodicity at the boundaries and how the resonance frequency of the modes changes from the mirror to the defect cell. The parameters used for the defects fabrication are the same as those expressed in Fig.~\ref{fig:nanobeam}~\textbf{a3} and were obtained from the optimization process.

The optical mode is confined at the X-point. The zero- and first-order modes (Fig.~\ref{fig:nanobeam}~\textbf{c1}) are both confined along the x-direction, parallel to the lattice parameter direction. While the first-order mode is confined mainly over the holes, where there is no material strain, the zero-order mode is distributed over the diamond, which tends to yield a higher value of the photoelastic component of the optomechanical coupling. For these modes, reducing the lattice parameter decreases the confinement region, increasing the resonance frequency. Thus defects were parameterized in a way that their lattice parameter decreases closer to the center of the nanobeam, creating a confining potential for the zero-order optical mode: $a_j=(1-d_j)a_N$, where the index $j$ refers to the number of the defect counting from the central defect ($j=0$) to the mirror defect $j=N$, $a_j$ is the $j$-th defect's lattice parameter, and $d_j$ is a function that softly varies from 1 to a defined value $d_0$. For the optical mode, the choice of $a_N$ is related to the frequency of the modes at the mirror, while $d_0$ determines how deeply the defect frequency penetrates in the bandgap. This amount can be increased or balanced by the variation of the hole area, which affects the effective cell refraction index. Increasing the hole reduces the refraction index of the defect which has the opposite effect of decreasing the lattice parameter.

The mechanical mode is confined at the $\Gamma$-point, where the zero-order mode (blue dots of Fig.~\ref{fig:nanobeam}~\textbf{b2}) becomes a simple displacement of the cell. The first mode (orange dots) is a ``breathing'' mode and normally is the mode that leads to the highest cooperativity since it can have high quality factor and optomechanical coupling with the zero-order optical mode. We have though observed less confined high-order modes with high-optomechanical coupling in the nanobeam simulation. The advantage of the breathing mode for confinement control is that its frequency strongly depends on the hole shape. If on one hand, the breathing mode has periodicity along the x-direction and therefore its frequency increases with the lattice parameter decreasing, on other hand, it is characterized by an increase of stress above the edge of the hole (Fig.~\ref{fig:nanobeam}~\textbf{b1}), thus squeezing the hole in the y-direction increases the stressed area decreasing the mode frequency. There is however a catch specific to this diamond nanobeam design that must be considered, that is the fact that for confining optical modes in the telecommunication C-band ($\sim\SI{200}{THz}$) the lattice parameter must be close to the $\SI{450}{nm}$ of thickness. This means that squeezing the holes in the x-direction may lead to modes with high enough energy to be confined only at the nanobeam surface, which may have a significant impact on the optomechanical coupling.

We opted to give the optimization method freedom to test different mechanical modes and confining strategies. The limitation was restricted only by the number of modes the simulation was allowed to find and by the symmetry constraints (only symmetric modes were simulated during the optimization process). The $x$ and $y$ components of the elliptical hole diameters were parameterized as $h_{xj} = (a_j-2\delta)\eta_j+\delta$ and $h_{yj} = (w-3\delta)\gamma_j+\delta$, where $\eta_j$ and $\gamma_j$ can vary between the defined parameters $\eta_0$, $\gamma_0$, $\eta_N$ and $\gamma_N$ (limited between 0 and 1) and $\delta=\SI{50}{nm}$, is a constant used to ensure that all simulated geometries are bounded by realistic fabrication limits. More details of the parameterization are given in the Supplement 1.

The high number of parameters to adjust, the way each of them influences at the same time the optical and mechanical modes, the different possible strategies for confinement, and the interplay between the mechanical confinement and the optomechanical coupling in diamond, make the nanobeam design optimization a complex problem with no trivial solutions.

\section{Nanobeam optimization}

To optimize the nanobeam design, we combined FEM simulations performed in COMSOL$^\text{\textregistered}$ with numerical optimization methods performed by Dakota$^\text{\textregistered}$. The simulations account for one-eighth of the nanobeam inside a computational domain (vacuum) bounded by PMLs, using proper mirror symmetry conditions to ensure that only symmetric mechanical modes are sought. The optical and mechanical studies are computed individually and used to evaluate the optomechanical coupling rate, which is composed of a photoelastic~\cite{Ramachandran1947} and a moving boundary component~\cite{Johnson2002}. The PML region ensures that radiative  optical and mechanical damping are accurately calculated (the PML typical absorption wavelength is changed according to the ongoing study).

Optimization methods that focus on maximizing the optomechanical coupling only, don't need to take losses into account. This simplifies a lot the optimization process because optimization is preserved if the nanobeam is rescaled. More than that, the optimization is preserved if small scaling is performed only in the $xy$-plane keeping the nanobeam thickness constant. This is usually done to adjust the design to the desired operating frequency. However, when dealing with the cooperativity, such scaling does not preserve the optimization, since losses do not scale together with the optomechanical coupling and the spectral distance between the defect modes and other bands which do not scale linearly with the geometry becomes a relevant factor. Consequently, the nanobeam must be optimized at a specific optical frequency. Therefore, we introduce an optical bandpass frequency filter $F(\nu_{o})=[\erf(\nu_{o}-\SI{183.9}{THz})-\erf(\nu_{o}-\SI{198.4}{THz})]/2$ when defining the optimization goal variable ($\erf$ is the error function), forcing solutions with optical resonance $\nu_o$ between a desirable frequency range, while leaving all geometric parameters free for optimization, except for the nanobeam thickness. Therefore, our design optimization consists on maximizing $q \equiv C_0 F(\nu_{o})$, where $C_0$ is the cooperativity of a single photon.

The optimization process is performed by Dakota$^\text{\textregistered}$ in two steps starting with a Genetic Algorithm (GA). In the second step, some solutions from the GA are selected and locally optimized using a Pattern Search (PS). Due to the complexity of the problem and the strong dependence on initial conditions, the combination of these two methods results in better solutions than the ones achievable by a single local optimization algorithm, in a short convergence time compared to global optimization algorithms. 

The GA is initially fed with random geometries containing a confined zero-order optical mode within the frequency range delimited by the frequency filter. This is ensured by a pre-start process where randomly generated geometries are tested. The geometries containing modes with a frequency above the frequency threshold are rescaled and reparemeterized. The ones without confined modes or with the zero-order mode frequency beyond the minimum threshold are discarded. For the rectangular nanobeam, approximately $45\%$ of the randomly generated geometries were used as the initial population of the GA and $70\%$ of them had to be rescaled.

To find the nanobeam zero-order optical mode frequency with a low computational cost, we simulate the defect unit cell with scattering bound conditions before the nanobeam study. The unit cell is much smaller and faster to simulate than the nanobeam, and its modes frequencies will be closer to the ones of the confined modes in the whole cavity, reducing the number of optical modes we have to simulate to find the zero-order mode.

The PS method was applied for 26 distinct solutions from the GA. The solutions were automatically selected by our algorithm, which excludes solutions too close to each other in the parameter space. The distance between two solutions were defined as $D \equiv \sum_{i} [(p_{1i}-p_{2i})/(p_{1i}+p_{2i})]^2$, where $p_{1i}$ and $p_{2i}$ refers to each one of the geometry parameters of the solutions. After discarding the solutions with the lowest $q$ value the algorithm compares pairs of solutions and, if they are too close to each other in the parameter space, the one with the lowest $q$ value is discarded. The threshold for $D$ to decide if a solution will be or not discarded is automatically varied until at least 20 solutions are selected.

By the end, the program returns the geometry parameters and all evaluated parameters (such as optomechanical coupling, quality factors, and modes frequency) for each tested solution.

\section{Results and Discussions}

\begin{figure*}[ht!]
\begin{center}
\includegraphics[width=12.5cm]{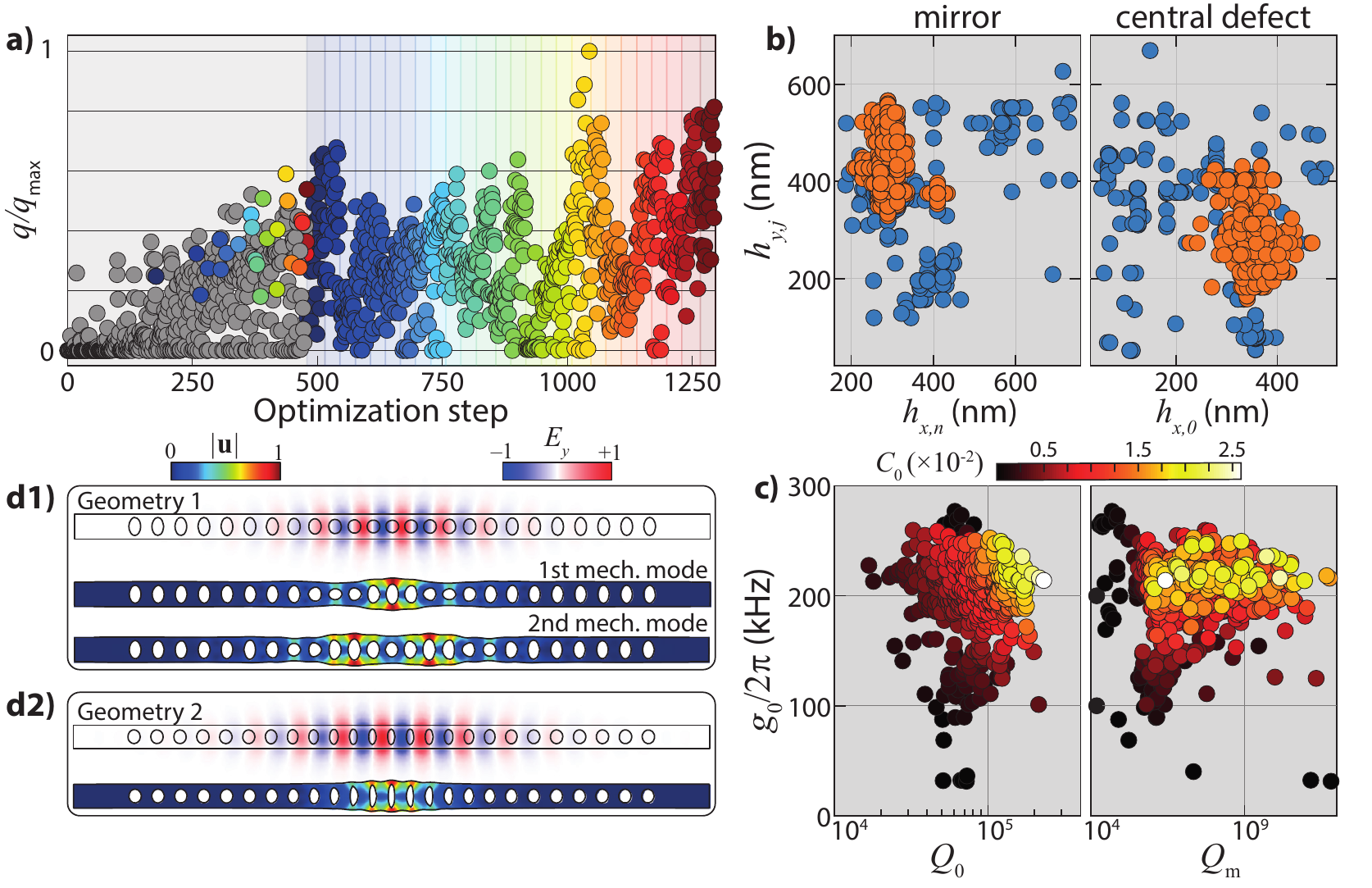}
\caption{\small{\textbf{a)} Value of the optimization parameter $q$ at each step of the optimization process. The first region (light gray region) are the 480 steps of the GA, the other 26 colors represent the PS applied for different initial parameters selected from the GA. The colored dots on the first region represent the distinct selected solutions from the GA for the PS. \textbf{b)} Hole diameters at the center ($j=0$) and mirror ($j=n$) for all tested geometry. The blue and orange dots refers to solutions of the GA and PS methods respectively. \textbf{c)} Optical (left) and mechanical (right) quality factor for each tested geometry. The color of the dots represent the value of the single photon cooperativity $C_0$. \textbf{d)} Ilustration of the optical and mechanical modes for two different geometries. The geometry represented in \textbf{d1} has two well-confined symmetric mechanical modes.}}
\label{fig:results}
\end{center}
\end{figure*}

\begin{table*}[ht!]
\centering
\begin{tabular}{c|c|c|c|c|c|c}
  & $\nu_o$~(THz) & $\nu_m$~(GHz) & $Q_o$ & $Q_m$ & $g_0/(2\pi)$~(kHz) & $C_0$ \\
  \hline
  1 & 194.28 & 8.37 & $2.3\times10^5$ & $\geq 10^6$ & 213.95 & $2.59\times10^{-2}$\\
  2 & 200.43 & 12.05 & $8.4\times10^4$ & $\geq 10^6$ & 306.74 & $1.31\times10^{-2}$\\
\end{tabular}
\caption{\small{Optical and mechanical frequencies ($\nu_o$ and $\nu_m$) and quality factors ($Q_o$ and $Q_m$), optomechanical coupling ($g_0$) and single-photon cooperativity ($C_0$) for two distinct solutions of the design optimization method. Both where optimized for $C_0 F(\nu_o)$. The first is the most optimized solution and the second is a particular solution using a distinct strategy to confine the mechanical mode.}}
\vspace{-6pt}
\label{tab:results}
\end{table*}

For the results presented in this paper we ran the GA with an initial population of 48 geometries and a limit of 480 evaluations. At the end, the PS was applied to 26 selected solutions with a minimum of 30 evaluations for each of them. We constrained an upper bound for both optical and mechanical quality factors of $1\times10^6$ to account for realistic material losses that ultimately limit these quantities. The graphics on Fig.~\ref{fig:results}~\textbf{a-c} summarize the results for all tested geometries. Fig.~\ref{fig:results}~\textbf{c} shows the optomechanical coupling, optical and mechanical quality factor and the cooperativity for the modes with the highest cooperativity in each tested geometry. While we were expecting solutions with the mechanical breathing mode confined between the zero- and first-order mirror modes, as shown in Fig.~\ref{fig:nanobeam}~\textbf{b2}, an unforeseen result was a solution with high cooperativity where the first-order mechanical mode was confined in a band formed between the first- and second-order modes. The Table~\ref{tab:results} contains the relevant properties of such solution compared with the solutions with highest $q=C_0 F(\nu_o)$. The most optimized solution presents two symmetric modes with high-cooperativity. The second mode is the one with the highest cooperativity since we limited the mechanical quality factor, and by being less localized, it has a larger overlap with the optical field and consequently a higher $g_0$. This mode has $\omega_m/\kappa = 43$, far in the resolved-sideband regime, and $C_0 = 2.59\times 10^{-2}$, implying that around 40 photons would be enough to reach the high-cooperativity regime, which is orders of magnitude bellow previously obtained experimentally in triangular diamond nanobeams~\cite{Burek2016}. Having two modes may be interesting from a fundamental point of view to study cooling in different regimes. The distance in frequency between these modes is around $\SI{228}{kHz}$ and depending on the mechanical quality factor of real devices they may overlap. There is also an antisymmetric mechanical mode for geometry 1 between the first and second presented modes that is not shown in Fig.~\ref{fig:results}~\textbf{d1}. We omitted the antisymmetryc mode because while it expands on one side of the nanobeam, it contracts in the opposite side. This makes the contribution to the $g_0$ from one half of the nanobeam to cancel with the other one, thus the total optomechanical coupling is negligible.

The solution for the second geometry has a lower optical quality factor, but a higher optomechanical coupling, which is interesting in a scenario where the material losses limit the optical quality factor. As for the first geometry, this solution can also reach the high-cooperativity regime at room temperature. Fig.~\ref{fig:results}~\textbf{d1} and \textbf{d2} show the geometry and the optical and mechanical modes for the most optimized (geometry 1) and the alternative (geometry 2) solutions, respectively. The difference between the first and second geometries resides in the way the hole shape changes.

\begin{figure}[ht!]
\begin{center}
\includegraphics[width=6.375cm]{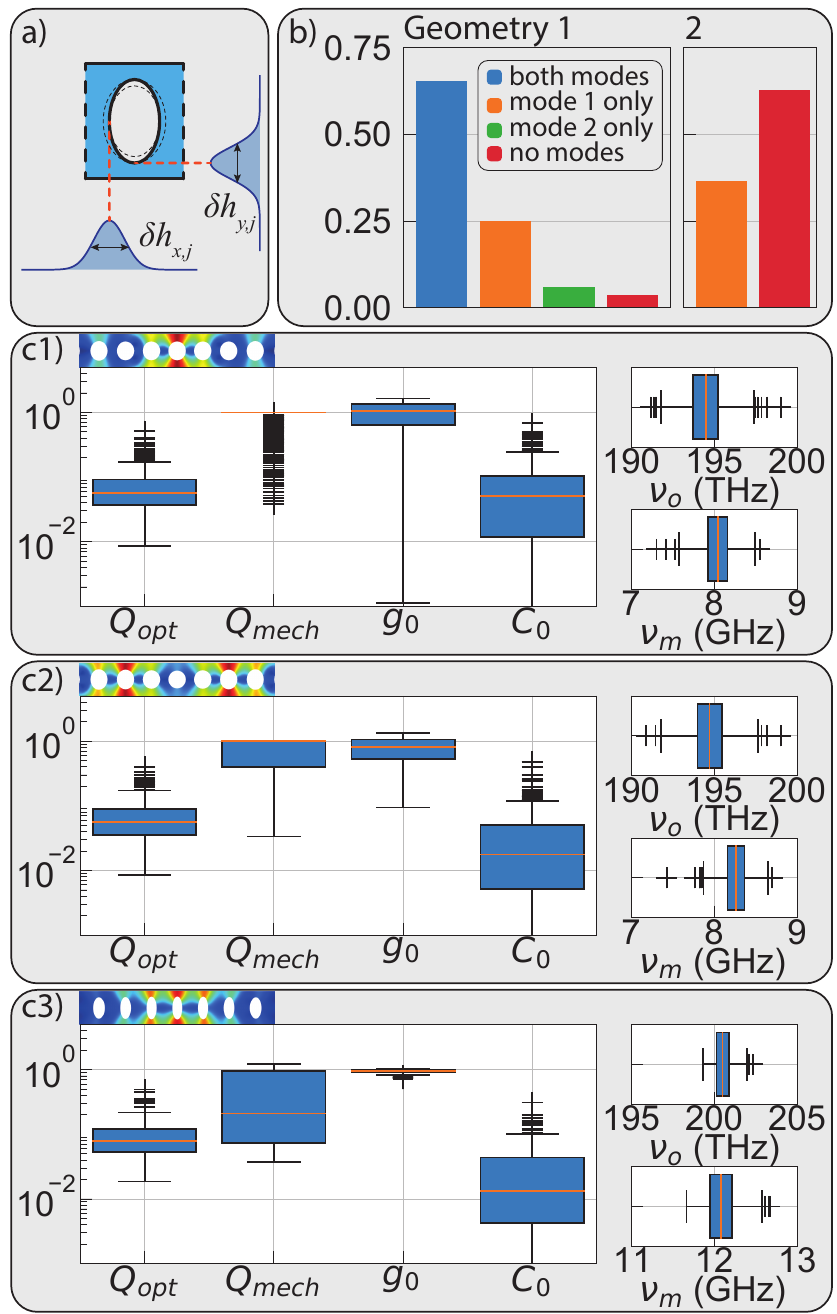}
\caption{\small{\textbf{a)} Illustration of the possible introduced fluctuations of the x and y components of the holes diameter. Each component of each hole is changed individually by a random amount that can be positive or negative, with normal distribution $N(\mu=0,\sigma = \SI{10}{nm})$.
\textbf{b)} Probability of successfully tracking well confined modes for geometries 1 and 2 according to the classification criteria which considers $p_{i,j}>0.75$ and $Q_m \geq 3\times10^4$ to confirm the right mode were tracked and it is still confined despite the introduced perturbations in the geometry. 
\textbf{c)} Statistical distribution of optical and mechanical quality factors, optomechanical coupling, and cooperativity normalized by the values of the ``perfect'' design (left) and the statistical distribution of the optical and mechanical frequencies (right).}}
\label{fig:resilience}
\end{center}
\end{figure}

At last, we performed a resilience test against imprecision of the size of the holes that are expected from the fabrication process. We introduce such imprecision by adding normal-distributed random value with null mean and standard deviation $\sigma = \SI{10}{nm}$ to $h_{xj},h_{yj}$ (Fig.~\ref{fig:resilience}~\textbf{a}). Due to the added imperfections, the mechanical modes change and it is important to define a way of tracking the breathing modes from Fig.~~\ref{fig:results}~\textbf{d}. We did this by computing, for each mode, the inner product of the displacement vectors from the ``perfect'' geometry with the ones of the ``imperfect'' geometry, at the nanobeam upper surface, which is the nanobeam surface that remains unchanged when adding imperfections to the holes. By normalizing this inner product we obtain the projection (at the upper surface) of a unknown mode over the modes we want to track:

\begin{equation}
    p_{i,j} = \frac{\langle m_i,m_j \rangle}{\|m_i\|\|m_j\|}
\end{equation}

We thus track the mechanical modes of the perturbed geometries by selecting the ones with the maximum projections. Since there are cases in which one or all modes we are tracking are heavily changed it is important to define thresholds for $p_{i,j}$ and $q$ where we assume these modes are not the same modes we had before the perturbation or they are not confined anymore. We used $p_{i,j}>0.75$ and $Q_m \geq 3\times10^4$ as our thresholds and then classified the perturbed geometries as having 2, 1 or no tracked modes (Fig.~\ref{fig:resilience}~\textbf{b}). For the first geometry, we successfully tracked both modes in most cases, although mode 1 (Fig.~\ref{fig:resilience}~\textbf{c1}) is more resilient. This is consequence of the fact that the first mechanical mode has simulated mechanical quality factor fair above $10^6$ and it remains above this value when introducing perturbations. For the second geometry, the mechanical quality factor is less resilient to perturbations due to the narrower bandgap, therefore, the mode confinement is weaker.

Concerning the cooperativity, for the first geometry it drops down around one order of magnitude on average for the first mechanical mode due to a drop in the optical quality factor (Fig.~\ref{fig:resilience}~\textbf{c1-2}) and a bit more for the second mechanical mode that experiments more losses, but the high-cooperativity regime is still reachable at room temperature for both modes at many simulated devices (from all simulated devices 44.2\% had $C_0>10^{-4}$). For the second geometry most of the tested devices did not match the tracking criteria (Fig.~\ref{fig:resilience}~\textbf{b}). For the devices in which we found the first-order confined mechanical mode, there is a drop down in the mechanical quality factor (Fig.~\ref{fig:resilience}~\textbf{c3}) probably due to the narrower mechanical bandgap, the lower mechanical confinement however makes $g_0$ very resilient and 62.0\% of these devices (23.6\% of all simulated devices) have $C_0>10^{-4}$. 

In summary, we have developed an efficient method to OMC cavities design optimization. The mixed search of our optimization scheme ensures a broad scan of the parameter space, leading to optimized solutions with distinct properties. For the rectangular nanobeam we investigated two solutions with single photon cooperativity $C_0 > 10^{-2}$ and distinct strategies to confine the mechanical modes. Resilience test suggests the high-cooperativity regime can be reached using both geometries. The developed optimization method is not limited to diamond devices neither to the rectangular nanobeam geometry.

\section*{Funding} 
This work was supported by S{\~a}o Paulo Research Foundation (FAPESP) through grants 2020/00100-9, 2020/00119-1, 2018/15580-6, 2018/15577-5, 2018/25339-4, 2021/10249-2, Coordena{\c c}\~ao de Aperfei{\c c}oamento de Pessoal de N{\'i}vel Superior - Brasil (CAPES) (Finance Code 001), Conselho Nacional de Desenvolvimento Cient{\'i}fico e Tecnol{\'o}gico, and Financiadora de Estudos e Projetos (Finep).

\section*{Disclosures} 
The authors declare no conflicts of interest.

\section*{Data availability}
Data underlying the results presented in this paper are available at ZENODO$^\text{\textregistered}$ repository (\href{https://doi.org/10.5281/zenodo.6560537}{10.5281/zenodo.6560537})~\cite{Dataset}, including FEM simulations, scripts files for generating figures and data related to the optimization method.

\section*{Supplemental document}
See Supplement 1 for supporting content. 


\end{document}


\maketitle

\section{Optimization process}

\subsection{Nanobeam parameterization}

The geometric parameters of the nanobeam are:

\begin{itemize}
    \item $w$ the nanobeam width,
    \item $t$ the nanobeam thickness,
    \item $N$ the number of defects from the central defect to the mirrors,
    \item $M$ the number of holes (defects + mirrors),
    \item $a_j$ the j-th cell's lattice parameter,
    \item $h_{xj}$ the j-th cell's x-component of the hole diameter,
    \item $h_{yj}$ the j-th cell's y-component of the hole diameter.
\end{itemize}

The four first parameters from the list were used as they are and the nanobeam width was the only one from them used as an optimization variable. The last three parameters were parameterized in function of:

\begin{itemize}
    \item $a_N$ the mirrors lattice parameter,
    \item $d_0$ to define the central defect's lattice parameter,
    \item $\gamma_N$ to define the x-component of the hole diameter at the mirror,
    \item $\gamma_0$ to define the x-component of the hole diameter at the central defect,
    \item $\eta_N$ to define the y-component of the hole diameter at the mirror,
    \item $\eta_0$ to define the y-component of the hole diameter at the central defect,
    \item $\delta = \SI{50}{nm}$ the fabrication precision limit.
\end{itemize}

To define the geometry of a unit cell between the mirrors and the central defect we define $x_j\equiv j/N$ where $j$ is the cell index counting from the central defect ($j=0$), to the first mirror ($j=8$) and a function $f(x_j)=2x_j^3-3x_j^2+1$, which is limited between 1 and 0 for our domain and has a null derivative at 0 and 1. The j-th lattice parameter is defined as:
\begin{align}
    a_j &= (1-d_j)a_N\label{eq:first}\\
    d_j &= d_0 f(x_j)
\end{align}

The equations to parameterize $h_{xj}$ and $h_{yj}$ are:
\begin{align}
    h_{xj} &= (a_j-2\delta)\eta_j+\delta\\
    h_{yj} &= (w-3\delta)\gamma_j+\delta\\
    \eta_j &= \eta_N-(\eta_N-\eta_0)f(x_j)\\
    \gamma_j &= \gamma_N-(\gamma_N-\gamma_0)f(x_j)\label{eq:last}
\end{align}
They ensure $\delta \leq h_{xj} \leq a_j$ and $\delta \leq h_{yj} \leq w - 2\delta$, which means no hole or wall is smaller than $\delta$. This parameterization requires $\gamma_N$, $\gamma_0$, $\eta_N$, $\eta_0$ and $d_0$ to be bounded between 0 and 1. 

\subsection{Simulation}

All simulations were performed in Comsol$^\text{\textregistered}$ using built-in diamond (100) and air as materials. Diamond has refractive index $n=2.417$ and Young's modulus given by $E=(C_{11}-C_{12})(C_{11}+2*C_{12})/(C_{11}+C_{12})=\SI{1050}{GPa}$. The elasticity matrix is defined as: 

\begin{equation*}
    D=
    \begin{pmatrix}
    C_{11} & C_{12} & C_{12} & 0 & 0 & 0\\
    C_{12} & C_{11} & C_{12} & 0 & 0 & 0\\
    C_{12} & C_{12} & C_{11} & 0 & 0 & 0\\
    0 & 0 & 0 & C_{44} & 0 & 0\\
    0 & 0 & 0 & 0 & C_{44} & 0\\
    0 & 0 & 0 & 0 & 0 & C_{44}
    \end{pmatrix}
\end{equation*}

with $C_{11}=\SI{1076}{GPa}$, $C_{12}=\SI{125}{GPa}$ and $C_{44}=\SI{578}{GPa}$. We also used the diamond photoelastic coefficients $(\rho_{11},\rho_{12},\rho_{44})=(-0.25,0.043,-0.172)$ to compute the photoelastic component of the optomechanical coupling.

The unit cell simulations account for a quarter of the cell using proper symmetry conditions at the internal boundaries, Floquet periodicity at the cell limits, and scattering boundary conditions (SBC) at the external boundaries as a filter for optical modes. The nanobeam simulations account for one eight of the nanobeam and were performed using perfectly matched layers (PML) instead of SBC at the external boundaries to improve accuracy on losses evaluation.

The geometries were created using the previously described parameterization through Comsol$^\text{\textregistered}$ methods. Optical and mechanical studies are computed independently by another Comsol$^\text{\textregistered}$ method, and the value of the integrals of optical fields are stored so they don't need to be recalculated during the evaluation of the optomechanical coupling for each mechanical mode, which strongly reduces each simulation's computational time. The simulation files are available at Zenodo$^\text{\textregistered}$ through the DOI:~\href{https://doi.org/10.5281/zenodo.6560537}{10.5281/zenodo.6560537}.

\subsection{Initial parameters generation}

The optimization method has two steps: a global optimization with a genetic algorithm (GA) and a multi-start local optimization using a pattern search (PS) method. To improve the efficiency of the GA we fed it with an initial population containing geometries capable of confining optical modes between $\SI{183.9}{THz}$ and $\SI{198.4}{THz}$. To generate such an initial population we first create a file with randomly generated geometries using Dakota LHS sampling with restricted conditions as described in Table~\ref{tab:pars_bounds}. Each generated geometry was tested by simulating first the central defect unit cell with periodic conditions at the transversal plane (yz) and scattering bound conditions at the other planes to identify the expected frequency for the nanobeam confined mode. After this we simulate the whole nanobeam and ran the optical study, looking for modes close to the expected value. At this point, the program discards all geometries that don't have optical modes with $Q_o>300$ and the geometries with low-frequency modes ($\nu_o<183.9$). The geometries with $\nu_o>198.4$ are then rescaled and reparameterized. Geometries with low frequency are not rescaled because they could make walls or holes to be thinner than the fabrication resolution limit.

\begin{table}[hb]
    \centering
    \begin{tabular}{c c c}
        parameter & lower bound & upper bound\\
        \hline
        $w$ & $\SI{400}{nm}$ & $\SI{800}{nm}$\\
        $a_N$ & $\SI{500}{nm}$ & $\SI{950}{nm}$\\
        $d_0$ & 0.05 & 0.6\\
        $\eta_N$ & 0.1 & 1\\
        $\gamma_N$ & 0.1 & 1\\
        $\eta_0$ & 0 & 1\\
        $\gamma_0$ & 0 & 1\\
    \end{tabular}
    \caption{Bounds restriction for GA initial population generation.}
    \label{tab:pars_bounds}
\end{table}

To scale the geometry to find a confined optical mode in the expected frequency range we estimate an expected scaling factor by dividing the frequency of the mode with the highest quality factor by the maximum allowed frequency and start to test scaling factors $S_f$ in a range between the expected factor more or less 10\%. We multiply all geometric parameters, except the nanobeam thickness and the number of holes, by the scaling factor and if a mode is found with an allowed frequency we reparameterize the geometry in a way the mirrors and the central defect cells are kept the same but the geometry can again be written by equations~\ref{eq:first}-\ref{eq:last}. The reparameterization equations are:

\begin{align}
    w' & = w S_f\\
    a' & = aS_f\\
    \eta_m' & = \frac{(a_N-2\delta)\eta_m S_f+\delta (S_f-1)}{a_N S_f-2 \delta}\\
    \eta_0' & = \frac{[a_N*(1-d_0)-2\delta]\eta_0 S_f+\delta (S_f-1)}{a_N*(1-d_0) S_f-2 \delta}\\
    \gamma_m' &= \frac{(w-3\delta)\gamma_m S_f+\delta (S_f-1)}{w S_f-3 \delta}\\
    \gamma_0' &= \frac{(w-3\delta)\gamma_0 S_f+\delta (S_f-1)}{w S_f-3 \delta}
\end{align}

\begin{figure*}[ht]
    \centering
    \includegraphics[width=12.5cm]{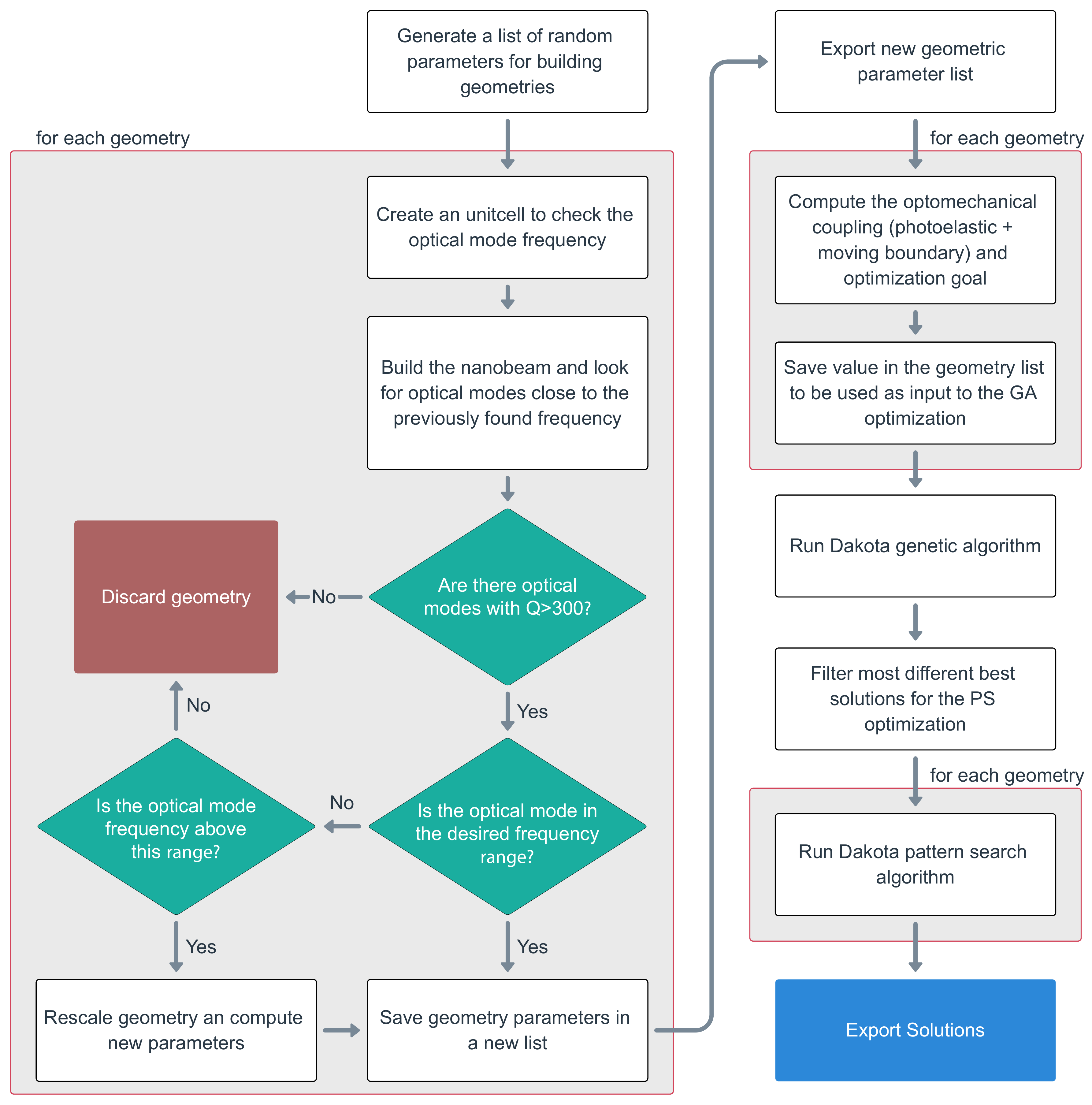}
    \caption{Flowchart of the optimization process.}
    \label{fig:flowchart}
\end{figure*}

After this, the new list containing the scaled geometries is used as input parameters for the GA. The full process is described in the flowchart of Fig.~\ref{fig:flowchart}.

\subsection{GA output selection}
The GA randomly mutates geometries looking for solutions with a higher value of the optimization goal $q$. Solutions with a low $q$-value are discarded and the remaining ones mutate again to generate new solutions. After a while, it is common though to reach a branch of solutions with high $q$ derived from the same solution and very similar to each other. Performing local optimizations in many solutions of the same branch is therefore an overkill process. Hence, to find solutions with different optimization strategies, we define clusters of solutions in the parameters space and select only the ones with the highest $q$. Two solutions are considered to belong to the same cluster if the normalized distance between then $D \equiv \sum_{i} [(p_{1i}-p_{2i})/(p_{1i}+p_{2i})]^2$, where $p_{1i}$ and $p_{2i}$ refers to each one of the geometry parameters of the solutions, is smaller than a minimum distance $D_{min}$. To define $D_{min}$ we initially discard all solutions with $q<0.3(q_{max}-q_{min})+q_{min}$, then we start the selection with $D_{min}=0$ and we increase $D_{min}$ by 0.03 until reach less than 20 selected solutions. At least we start to slowly decrease $D_{min}$ again by 0.001 until reach at least 20 selected solutions.

\section{Output geometries}
The parameter values for the geometries presented in the paper are listed in Table~\ref{tab:build_pars}

\begin{table}[hb]
    \centering
    \begin{tabular}{c c c}
        parameter & geometry 1 & geometry 2\\
        \hline
        $N$ & $8$ & $8$\\
        $M$ & $13$ & $13$\\
        $t$ & $\SI{450}{nm}$ & $\SI{450}{nm}$\\
        $w$ & $\SI{633.9}{nm}$ & $\SI{584.3}{nm}$\\
        $a_N$ & $\SI{610.8}{nm}$ & $\SI{569.2}{nm}$\\
        $d_0$ & 0.184 & 0.178\\
        $\eta_N$ & 0.513 & 0.557\\
        $\gamma_N$ & 0.839 & 0.677\\
        $\eta_0$ & 0.733 & 0.357\\
        $\gamma_0$ & 0.561 & 0.773\\
    \end{tabular}
    \caption{Parameters of the optimized solutions presented in the paper.}
    \label{tab:build_pars}
\end{table}

\section*{Acknowledgement}
This work was supported by S{\~a}o Paulo Research Foundation (FAPESP) through grants 2020/00100-9, 2020/00119-1, 2021/10249-2, 2018/15580-6, 2018/15577-5, 2018/25339-4, Coordena{\c c}\~ao de Aperfei{\c c}oamento de Pessoal de N{\'i}vel Superior - Brasil (CAPES) (Finance Code 001), Conselho Nacional de Desenvolvimento Cient{\'i}fico e Tecnol{\'o}gico, and Financiadora de Estudos e Projetos (Finep).